\documentstyle[prb,aps]{revtex}

\begin{document}
\draft
\title{Phase diagram of a stochastic cellular automaton with long-range interactions}

\author{Sergio A. Cannas\cite{auth}}  

\address{Facultad de Matem\'atica, Astronom\'\i a y F\'\i sica, 
Universidad Nacional de C\'ordoba,  Ciudad Universitaria, 5000 C\'ordoba, Argentina
\cite{auth3} }

\date{\today}
\maketitle

\begin{abstract} 
 A stochastic one-dimensional cellular automaton with long range spatial interactions is introduced. In this model the probability state at time $t$ of a given site depends on the state of all the other sites at time $t-1$ through a power law of the type $1/r^\alpha$, $r$ being the distance between sites. For $\alpha\rightarrow\infty$ this model reduces to the Domany-Kinzel cellular automaton. The dynamical phase diagram is analyzed using Monte Carlo simulations for $0\leq\alpha\leq\infty$. We found the existence of two different regimes: one for $0\leq\alpha\leq 1$ and the other for $\alpha>1$. It is shown that in the first regime the phase diagram becomes independent of $\alpha$. Regarding the frozen-active phase transition in this regime, a strong evidence is found that the mean field prediction for this model becomes exact, a result already encountered in magnetic systems. It is also shown that, for replicas evolving under the same noise, the long-range interactions fully suppress the spreading of damage for  $0\leq\alpha\leq 1$.
\end{abstract}

\section{Introduction}

The Domany-Kinzel cellular automaton\cite{Domany} (DK herein) has been vastly studied in the last years. Its combined simplicity and rich dynamical behavior make it a prototype for studying and testing different theoretical aspects of complex systems, such as dynamical critical properties\cite{Martins2}, directed percolation\cite{Domany,Grassberger} and spreading of damage\cite{Martins,Hinrichsen}  properties.

The DK model consists in a one dimensional lattice with binary variables associated to each site $i$ ($S_i=0$: ``dry'' or ``death'' state and $S_i=1$: ``active'' state). The system evolves at discrete time, with the state of each site depending stochastically on the configuration of its two nearest neighbor sites at the previous time, independent of the other sites at the same time. In other words, the interactions in the DK model are local or short ranged both in space and time. Then, the dynamics of the model is defined by four independent parameters, namely, the transition probabilities corresponding to the four different configurations of the nearest neighbors of any site.

A simple version of the model is obtained by equaling to zero the probability of any site to be in the active state given that its neighbors were both in the death state (which defines a so-called ``legal'' rule of evolution), and considering only one value for the probability of the site to be in the active state given that only one of its neighbors were in the active state, independent of which one. This leads to a two-parameter model which contains most of the complex behavior of the model. Domany and Kinzel showed that this model presents a continuous transition between a ``frozen'' phase where all the sites are in the death state (this is an absorbing state from which the system cannot scape once it is reached) and an ``active'' phase with a constant average fraction of sites in the active state. Some years after Martins {\it et al} showed\cite{Martins}, using the spreading of damage technique, that the active phase is divided  two regions: one in which the system presents sensitivity to the initial conditions and one in which it does not. Such regions were interpreted as dynamical phases and the transition between them was shown to be a continuous one. The sensibility to initial conditions is studied by looking at the time evolution of the Hamming distance between two replicas of the system ({\it i.e.}, two identical systems with different initial configurations) under correlated noise\cite{Martins}. Recently, Hinrichsen {\it et al} showed that the active phase is actually subdivided into three dynamical phases which depend on the correlations between the different types of stochastic noise used for each replica. The phases are the following: ({\it i}) one in which the damage spreads for all possible types of correlations, which coincides with the so called chaotic phase founded by Martins {\it et al}; ({\it ii}) one in which the damage spreads  for some type of noise but it does not others and ({\it iii}) one where damage heals always.

The  phase diagram of this model was calculated using different techniques, such as mean field\cite{Tome,Kohring,Bagnoli} and Monte Carlo\cite{Martins2,Hinrichsen,Zebende}. 

In this work we introduce a generalization of the above model, in which the state of any site depends on the state of {\bf all} the rest of the sites at the previous time through a distance dependent transition probability. In this case the interaction between a pair of sites separated by a distance $r$ decays with a power law of the type $1/r^\alpha$ ($\alpha\geq 0$). In other words, the interactions are long ranged in space. Microscopic long-range interactions appear in different physical contexts (see Ref.\cite{Cannas} and references therein) but, until now, most of the theoretical analysis has been concentrated into the equilibrium properties of such kind of systems. Hence, it is of interest to analize systematically the influence of long range interactions in a dynamical model, such as a cellular automaton.

\section{The model: long-range cellular automaton}

Let us consider a stochastic one-dimensional lattice with $N+1$ sites. The configurations of the system are described in terms of a set of occupation variables $S_i=0,1$ with $i=0,1,\ldots,N$. Let $S^t_i$ the state of the site $i$ at the discrete time $t$. We denote by $S^t\equiv \left\{S^t_i\right\}$ the state of the complete system at time $t$. We consider periodic boundary conditions, so that

\begin{equation}
S^t_{i+N+1} = S^t_i \;\;\;\;\;\;\;\; i=0,1,\ldots,N
\label{pbc}
\end{equation}

The system evolves in time according to a spatially {\bf global} rule, but without memory effects, that means, the state of the site $i$ at time $t$ depends on the state of {\bf all} the rest of the sites $S^{t-1}$ at the previous time $t-1$. However, the influence of a site $j$ at time $t-1$ over the site $i$ at time $t$ decays with the distance between sites with a power law $\left| i-j \right|^\alpha$ with $\alpha\geq 0$. Before establishing explicitly the evolution rule we introduce an auxiliary function $p'(S_3|S_1,S_2)$ of three binary variables $S_1,S_2$ and $S_3$, such that

\begin{equation}
p'(0|S_1,S_2)+p'(1|S_1,S_2)=1\;\;\;\;\;\;\;\forall S_1,S_2
\label{eq2}
\end{equation}

\noindent and

\begin{eqnarray}
p'(1|0,0) &=& 0 \nonumber\\
p'(1|1,0) &=& p'(1|0,1) = p_1 \label{eq2'}\\
p'(1|1,1) &=& p_2 \nonumber
\end{eqnarray}

\noindent with

\begin{equation}
0 \leq p_1,p_2 \leq 1.
\label{eq4}
\end{equation}

\noindent Notice that $p'(S_3|S_1,S_2)$ has the properties of a conditional probability for the variable $S_3$ given the values of $S_1$ and $S_2$. Now, we consider a parallel dynamics, so that all the sites are updated simultaneously and independent of the rest of the sites at the same time. Then, the transition probability $W(S^t|S^{t-1})$ from state $S^{t-1}$ to $S^t$ is given by

\begin{equation}
W(S^t|S^{t-1})= \prod_{i=0}^N w(S^t_i|S^{t-1})
\label{eq5}
\end{equation}

\noindent where $w(S^t_i|S^{t-1})$ is the probability that the site $i$, at time $t$, has the value $S^t_i$ given that, at time $t-1$, the system was at $S^{t-1}$.

The model is then defined by the expression:

\begin{equation}
w(S^t_i|S^{t-1})=\frac{\sum_{n=1}^{N/2} \frac{1}{n^\alpha}\; 
                  p'(S^t_i|S^{t-1}_{i-n},S^{t-1}_{i+n})}
                  {\sum_{n=1}^{N/2} \frac{1}{n^\alpha}}
\label{modelo}
\end{equation}

\noindent where $N$ is assumed an even number for simplicity and $\alpha\geq 0$. Note that the interactions are symmetric on both sides of site $i$ and that, together with the periodic boundary conditions (\ref{pbc}) every interaction is counted just once on a ring of length $N+1$, as schematized in Fig.1. The sum $\sum_{n=1}^{N/2} \frac{1}{n^\alpha}$ that appears in the denominator of Eq.(\ref{modelo}) ensures the correct normalization of the conditional probability $w(S^t_i|S^{t-1})$. It is also related\cite{Cannas1} to some function $N^*(\alpha)$ that provides the correct scaling of the thermodynamic functions of magnetic models\cite{Tsallis} for $\alpha\leq 1$, where these kind of interactions are non integrable.

From Eqs.(\ref{eq2})-(\ref{eq4}) is easy to verify the following properties:

\begin{equation}
0\leq w(S^t_i|S^{t-1})\leq 1
\end{equation}

\begin{equation}
\sum_{S_i=0,1} w(S_i|S^{t-1})=1  \;\;\;\;\;\forall S^{t-1}
\label{normalization}
\end{equation}

\noindent and from Eq.(\ref{eq5}):

\begin{equation}
\sum_{S^t} W(S^t|S^{t-1})=1  \;\;\;\;\;\forall S^{t-1}
\end{equation}

In the limit $\alpha\rightarrow\infty$ we have that

\begin{equation}
\lim_{\alpha\rightarrow\infty} w(S^t_i|S^{t-1})=
          p'(S^t_i|S^{t-1}_{i-1},S^{t-1}_{i+1})
\end{equation}

\noindent and the model reduces to the DK cellular automaton \cite{Domany}. For finite values of $\alpha$ the transition probability for the site $i$ depends on all the rest of the sites. For small values of $\alpha$ the interaction strength tends to become independent of $\alpha$. This is exactly the case for $\alpha=0$ where the system lost its dimensionality. In the limit $N\rightarrow\infty$ this model ($\alpha=0$) can also be thought as an infinite dimensional one. In magnetic systems the mean field approximation becomes exact in the limit of infinite dimensionality. Moreover, in a recent study of the one-dimensional Ising model with long-range interactions of the type $1/r^\alpha$ it has been shown that the mean field behaviour is exact, not only in the infinite dimensional case $\alpha=0$ (which corresponds to the Curie-Weiss model) but in the full range of interactions\cite{Cannas1} $0\leq \alpha\leq 1$. This result suggests that mean field behaviour could be an universal property of systems with long-range interactions, from which the infinite dimensional case $\alpha=0$ is just a particular case. Therefore, it is interesting to investigate whether or not this property is shared by dynamical models with long-range interactions, like the present one.

In a very broad sense (and perhaps a rather vague one) a mean field approximation is one in which fluctuations are neglected.  Different systematic mean field approaches has been proposed for the DK cellular automaton\cite{Tome,Bagnoli,Kohring}. They all coincide (at least qualitatively) in the prediction for the shape of the frozen-active transition line, but they differ in the prediction of the damage spreading transition line. Hence, it is of interest to analize the behaviour of the long-range cellular automaton for small values of $\alpha$, where some kind of mean-field type behaviour can be expected, by analogy with the magnetic models.

Let $P_t(S)$ be the probability of state $S$ at time $t$. The time evolution of this function is given  by the equation:

\begin{equation}
P_{t+1}(S)=\sum_{S'} W(S|S') P_t(S').
\label{Pt}
\end{equation}

We are mainly interested in the asymptotic behaviour of this model for $t\rightarrow\infty$. Since this behaviour is not expected to depend on the initial state (except for some very special cases, for instance, $S^t_i=0$ $\forall i$, where the solution is trivial), we consider, without loss of generality, the following initial state:

\begin{equation}
P_0(S) = \prod_{i=0}^N \frac{1}{2} \left(\delta_{S_i,0}+\delta_{S_i,1}\right)
\end{equation}

\noindent where $\delta_{S_i,S_j}$ is a Kronecker delta function.

We denote by $\left< f(S) \right>_t$ the mean value of a state function $f(S)$ at time $t$:

\begin{equation}
\left< f(S) \right>_t\equiv \sum_S f(S) P_t(S)
\end{equation}

The mean activity is then defined as

\begin{equation}
a_N(t) \equiv \frac{1}{N+1} \left< \sum_{i=0}^N S_i \right>_t =
        \frac{1}{N+1} \sum_{i=0}^N S_i P_t(S_i)
\end{equation}

\noindent where $P_t(S_i)$ is the marginal probability for $S_i$:

\begin{equation}
P_t(S_i) \equiv \sum_{\left\{ S_{j\neq i} \right\}} P_t(S).
\label{marginal}
\end{equation}

Since we are considering an homogeneous system $P_t(S_i)$ is independent of the site $i$. Then, defining $P_t(1)\equiv P_t(S_i=1)$ we have that

\begin{equation}
a_N(t) = P_t(1)
\label{at}
\end{equation}

This kind of system is expected to present two type of homogeneous stationary states  for $t\rightarrow\infty$: one with $S_i=0$ $\forall i$, which defines the so-called ``frozen'' or ``dry'' phase, and one with a finite fraction of active sites $S_i=1$, which defines the ``active'' phase. The order parameter for this phase transition is given by

\begin{equation}
a_N \equiv \lim_{t\rightarrow\infty} a_N(t).
\end{equation}

From Eqs.(\ref{eq5}), (\ref{Pt}) and (\ref{at}) the time evolution of $a_N(t)$ is given by the equation:

\begin{equation}
a_N(t+1) = \left< w(S_i=1|S')\right>'_t
\label{at2}
\end{equation}

\noindent for an arbitrary site $i$. Let $P_t(S_i,S_j)$ the joint marginal probability for $S_i$ and $S_j$. From Eqs.(\ref{eq2'}), (\ref{modelo}) and (\ref{at2}) we have that

\begin{equation}
a_N(t+1) = 2\, p_1\, a_N(t) + \frac{p_2-2\, p_1}{\sum_{n=1}^{N/2}                                                \frac{1}{n^\alpha}}\;
           \sum_{n=1}^{N/2} \frac{1}{n^\alpha} P_t^n(1,1)
\label{at3}
\end{equation}

\noindent where 

\[ P_t^n(1,1) \equiv P_t(S_{i-n}=1,S_{i+n}=1) \]

\noindent and we have used that 

\[ P_t(1,0) = P_t(0,1) = P_t(1)-P_t(1,1)  \]

\noindent for all pair of sites $i\neq j$.

\section{The frozen-active phase transition}

Before proceeding with the mean-field and Monte Carlo calculations we can obtain a few analytical properties of the phase diagram in the $(p_1,p_2)$ space, regarding the frozen-active phase transition.

First of all, notice that, from Eqs.(\ref{eq2'}) and (\ref{modelo}), the absorbing state $a_N=0$ ($S_i=0\;\forall i$) is always a fixed point of the dynamics for all values of $\alpha$ and $N$. That means, $a_N(t)=0$ implies that $a_N(t+1)=0$. This state is expected to be an attractor of the dynamics for low values of $p_1$, as occurs in the DK automaton ($\alpha=\infty$).

On the other hand, for the particular case $p_2=1$ the fully occupied state $a_N=1$ ($S_i=1\;\forall i$) is also a fixed point of the dynamics for all values of $\alpha$ and $N$. If we now perform a ``particle-hole'' transformation, by defining a new set of variables $S'_i=1-S_i$ $\forall i$, we obtain, from Eqs.(\ref{eq2})-(\ref{modelo}) and Eq.(\ref{normalization}), an equivalent system with parameters $p'_2=1$ and $p'_1=1-p_1$. We also see that $a'_N=1-a_N$. Therefore, if the system presents a frozen-active phase transition for $p_2=1$ then the critical value of $p_1$ has to be $p_1=p'_1=1/2$. Moreover, in the active phase $a_N=1$ and the phase transition will be a discontinuous one. These results will be verified by both the mean-field and Monte Carlo calculations.

\subsection{Mean-field solution}

From Eq.(\ref{at3}) we see that the time evolution of $a_N(t)$ depends on higher order correlation functions between sites. A simple dynamical mean-field approximation can be obtained by neglecting such correlations \cite{Tome}, that is, 

\begin{equation}
P_t(S_i,S_j)=P_t(S_i)P_t(S_j)\;\;\;\;\;\;\forall\; i,j
\end{equation}

\noindent which implies that

\begin{equation}
P^n_t(1,1) = \left[ P_t(1) \right]^2 = a_N^2(t)
\end{equation}

Replacing into Eq.(\ref{at3}) we find that

\begin{equation}
a_N(t+1) = 2\, p_1\, a_N(t) + (p_2-2\, p_1) a_N^2(t)
\label{at4}
\end{equation}

This equation is independent of $\alpha$ and $N$, and it is the same already encountered (with the same approximation) for the DK automaton\cite{Tome}. The fixed point equation $a_N(t+1)=a_N(t)=a_N$ derived from Eq.(\ref{at4}) gives the critical line $p_1^c=1/2$ $\forall p_2$ with

\begin{equation}
a_N= \frac{2\, p_1-1}{2\, p_1-p_2}
\label{MF}
\end{equation}

\noindent in the active phase $p_1\geq p_1^c$.

We can now develop a systematic improvement of this mean-field approximation, by neglecting the pair correlations only between sites located at a distance greater than K of site $i$, for some integer value $K$, that is, 

\begin{equation}
P^n_t(1,1) = \left[ P_t(1) \right]^2 \text{ for } n>K
\end{equation}

\noindent with $1\leq K<N/2$. Then, the Eq.(\ref{at4}) can be rewritten as

\begin{equation}
a_N(t+1) = 2\, p_1\, a_N(t) + (p_2-2\, p_1) a_N^2(t)
        \left( 1-\frac{\sum_{n=1}^{K} \frac{1}{n^\alpha}}
                      {\sum_{n=1}^{N/2} \frac{1}{n^\alpha}}\right)+
           \frac{p_2-2\, p_1}{\sum_{n=1}^{N/2}                                                \frac{1}{n^\alpha}}\;
           \sum_{n=1}^{K} \frac{1}{n^\alpha} P_t^n(1,1)
\label{at5}
\end{equation}

For large lattice sizes $N \gg 1$ the sum $\sum_{n=1}^{N/2}\frac{1}{n^\alpha}$ diverges for $0\leq\alpha\leq 1$, while

\begin{equation}
\sum_{n=1}^{N/2}\frac{1}{n^\alpha} \sim \zeta(\alpha) \text{ for } \alpha>1,
\end{equation}

\noindent where $\zeta(\alpha)$ is the Riemann Zeta function. Hence, for any {\it finite} value of $K$, Eq.(\ref{at5}) reduces to Eq.(\ref{at4}) in the limit $N\rightarrow\infty$, for $0\leq\alpha\leq 1$. This fact suggests that the mean field solution becomes exact for an infinite lattice when  $0\leq\alpha\leq 1$, while a departure from such solution should be expected for $\alpha>1$, a behaviour that has  already been observed in the ferromagnetic Ising model with long-range interactions, as was mentioned in the preceeding section. In fact, the Monte Carlo simulations of the next subsection confirm such conjecture.

\subsection{Monte Carlo results}

We performed a Monte Carlo simulation for several system sizes, $N=100$, $200$, $400$ and $800$, and for different values of $\alpha$. It is worth mentioning that the computational complexity of the algorithm implied by the long-ranged transition probability (\ref{modelo}) is $O(N^2)$.

We calculated the transition line between the frozen and active phases, founding a continuous transition for all values of $\alpha$ in the full plane $(p_1,p_2)$, except for $p_2=1$, where the numerical results verify the prediction of the beginning of this section. That is, for $p_2=1$, the transition point always occurs at $p_1^c=1/2$ for all values of $N$ and $\alpha$, and the transition is a discontinuous one at that point. The typical shapes of the transition lines between the frozen (left part of the diagram) and active (right part of the diagram) phases are shown in Fig.2 for $N=800$ and different values of $\alpha$. The curves show all the same qualitative shape as the corresponding one for the short-range model $\alpha=\infty$ (DK model). We can observe that, for $\alpha$ running from $\alpha=\infty$ to $\alpha=1$ the curves move smoothly to the left of the diagram. All the curves with $\alpha$ between $0$ and $1$ are indistinguishable inside the error intervals of the calculation (the error bars in the $p_1$ direction are approximately of the order of the symbol size; there are no error bars in the $p_2$ direction). This behaviour was verified for all the values of $N$ used in the simulation.

For each checked value of $\alpha>1$ the corresponding curves show a fast convergence into an asymptotic one for increasingly higher values of $N$. On the other hand, for $0\leq\alpha\leq 1$ the curves show a slow convergence towards the $p_1=1/2$ vertical line, as depicted in Fig.3 for $\alpha=0$. In fact, we verified for several values of $p_2$ the asymptotic behaviour $p_1^c-1/2 \sim A(p_2)\, N^{-1/2}$ for high values of $N$, where $A(p_2)$ is some continuous function of $p_2$ with $A(1)=0$. The Fig.4 illustrates the typical behaviour of $p_1^c$ as a function of $\alpha$, for several values of $N$ and a particular value of $p_2$. The circles in that figure correspond to a numerical extrapolation for $N\rightarrow\infty$.

In Fig.5 we show a numerical extrapolation for $N\rightarrow\infty$ of the full critical lines for different values of $\alpha$.

Finally, in Fig.6 we show a numerical extrapolation for $N\rightarrow\infty$ of the mean activity $a = lim_{N\rightarrow\infty} a_N$ {\it vs} $p_1$ curves (symbols) for three different values of $p_2$ and $0\leq\alpha\leq 1$ (the activity curves for different values of $\alpha$ in this range are indistinguishable). The solid lines are the mean field prediction of Eq.(\ref{MF}). The excellent agreement between these results confirm the conjecture of the preceeding subsection.

\section{Spreading of Damage}

To study the spreading of damage in this model we have to create  a replica of the system with an initial ``damage'', by flipping randomly a fraction $p$ of sites\cite{Martins}. Then we let both replicas to evolve under correlated noise and analize the long time evolution of the normalized Hamming distance between replicas. If the Hamming distance evolves to zero it is said that the damage ``heals''; otherwise, the damage ``spreads''.

In a recent work, Hinrichsen\cite{Hinrichsen} {\it et al} have shown that the damage can spread in the DK automaton for different kinds of correlated noise. In this work we let the replicas evolve under the {\bf identical realizations} of the stochastic noise, which was the most studied case in the DK automaton\cite{Martins,Tome,Zebende,Bagnoli,Grassberger}. This corresponds, according to Hinrichsen {\it et al} definition, to maximally correlated noise, at least for the DK automaton\cite{Hinrichsen} ($\alpha=\infty$). It is expected this property to be valid for all $\alpha$, as long as in this case a single random number is generated for every site in both replicas. Other types of correlated noise are very difficult to simulate in the long-range model, because they involve, in principle, the generation of $O(2^N)$ random numbers for every site at every time step.

We performed this Monte Carlo simulation for $N=100$, $200$ and $400$ founding that, for $0\leq\alpha\leq 1$, the damage spreading is suppressed for all values of $(p_1,p_2)$. For $\alpha>1$ a small region  where the damage spreads appears in the lower right corner of the diagram. The transition line moves continuously  towards the DK one for $\alpha=\infty$.

In Fig.7 we show  the transition lines between the healed and the ``chaotic'' regions, for different values of $\alpha>1$ and $N=400$.

\section{Conclusions}

We introduced a one-dimensional stochastic cellular automaton which allows to analyze different ranges of interactions through a single parameter $\alpha$. We verified that this model reproduces, for high values of $\alpha$ most of the known results of the DK model ($\alpha=\infty$).

We found two different regimes, according to the range of values of $\alpha$.

For $\alpha>1$ the dynamical phase diagram shows the same qualitative aspects of the one corresponding to the DK model, that is, three different regions: a frozen and an active phase, with the last one divided in two regions, one where the damage spreads (``chaotic'' phase) and one where the damage heals, when the replicas evolve under the same noise. The transition lines between these regions continuously vary when $\alpha$ decreases from $\alpha=\infty$ to $\alpha=1$.

For $0\leq\alpha\leq 1$ the phase diagram becomes independent of $\alpha$, even for finite $N$. No damage spreading were found in this case, at least for replicas evolving under the same noise. Regarding the frozen-active phase transition, we presented a strong evidence that the mean field theory is {\bf exact} in the limit $N\rightarrow\infty$. The fact that the same behaviour appears in the thermodynamics of  the one-dimensional Ising model\cite{Cannas1} with similar  interactions\cite{nota2} may suggests the existence of some {\bf universal} relationship between mean field theories and very long ranged interactions ($0\leq\alpha\leq d$, where $d$ is the space dimension). 

It is worth noting that mean field calculations of the Hamming distance dynamics between replicas evolving under two different types of correlated noise\cite{Tome}-\cite{Bagnoli} predicts  a spreading of damage transition in the DK model. It is a general heuristic argument (based on the equilibrium behaviour of magnetic systems) that mean field approximations of short range models ($\alpha=\infty$) show the same behaviour as the corresponding fully connected models ($\alpha=0$). We have shown that this is not true for the spreading of damage transition in the DK model. This variance between the two phenomena in the  cellular automata is telling us that the spreading of damage transition is of different nature, at least in some of its properties, than other cooperative phenomena like the frozen-active or the ferromagnetic phase ones.  

Finally, there is  recent conjecture of Grassberger\cite{Grassberger} which states that all continuous transitions from an ``absorbing'' state to an ``active'' one with a single order parameter are in the universality class of directed percolation, provided that some conditions are fulfilled, among others, {\bf short-range interactions} in space and time. Both transitions in the present model fall into this category, except for the range of the interactions for small values of $\alpha$. Hence, having a model which allows to interpolate between long and short range regimes, and for which the long range behaviour is known exactly may be a good candidate for testing the above conjecture.

I am indebted to  Constantino Tsallis for many fruitful
suggestions and discussions about this problem. Useful discussions with M. L. Martins, T. J. Penna and M. de Oliveira are acknowledged. I also 
acknowledge warm hospitality received at the Centro Brasileiro
de Pesquisas F\'\i sicas (Brazil), where this work was partly
carried out.  This work was partially supported by grants
from Funda\c{c}\~ao Vitae (Brazil), 
Consejo Nacional de Investigaciones Cient\'\i ficas y T\'ecnicas
CONICET (Argentina), Consejo Provincial de
Investigaciones Cient\'\i ficas y Tecnol\'ogicas CONICOR (C\'ordoba,
Argentina) and  Secretar\'\i a de Ciencia y
Tecnolog\'\i a de la Universidad Nacional de C\'ordoba
(Argentina).

\begin{figure}
\caption{Schematic diagram of interactions for the long-range cellular automaton in a ring of $N+1$ sites with periodic boundary conditions. The probability of site $i$ to be in a given state at time $t$ depends sym metrically on  all the rest of the sites along the ring (at time $t-1$), down to the site $i-N/2$ and up to the site $i+N/2$ ($N$ is assumed an even number).}
\label{fig1}
\end{figure}

\begin{figure}
\caption{Monte Carlo calculation of the transition lines between the frozen (left part of the diagram) and the active (right part) phases for $N=800$ and different values of $\alpha$. The lines are a guide to the eye.}
\label{fig2}
\end{figure}

\begin{figure}
\caption{Monte Carlo calculation of the transition lines between the frozen (left part of the diagram) and the active (right part) phases for $\alpha=0$ and different values of $N$. The lines are a guide to the eye.}
\label{fig3}
\end{figure}

\begin{figure}
\caption{Monte Carlo calculation of the critical value $p_1^c$ as a function of $\alpha$ for $p_2=0.5$ and different values of $N$. The circles correspond to a numerical extrapolation for $N\rightarrow\infty$. The error bars, when not shown, are smaller or equal than the symbol size. The lines are a guide to the eye.}
\label{fig4}
\end{figure}

\begin{figure}
\caption{Numerical extrapolation for $N\rightarrow\infty$ of the critical lines for different values of $\alpha$.}
\label{fig5}
\end{figure}

\begin{figure}
\caption{Numerical extrapolation for $N\rightarrow\infty$ of the mean activity $a$ (symbols) compared with the mean field prediction (solid lines) for $0\leq\alpha\leq 1$ and different values of $p_2$.}
\label{fig6}
\end{figure}

\begin{figure}
\caption{Transition lines between the healed (left part of the diagram) and the ``chaotic'' (right part) phases, for different values of $\alpha>1$ and $N=400$.}
\label{fig7}
\end{figure}


\begin{references}
\bibitem{nota2} There is not a simple mapping for finite $\alpha$ between this model and the equilibrium thermodynamics of a magnetic model, as occurs in the DK model\cite{Domany}.
\end{references}
\end{document}